%% file: main.tex
\documentclass[12pt]{article}
\usepackage[utf8]{inputenc}
\usepackage{graphicx,psfrag,epsf}
\usepackage{booktabs}
\usepackage{textgreek}
\usepackage{threeparttable}
\usepackage{float}
\usepackage{amsmath,amsfonts,amsthm,bm} 
\usepackage{url}
\usepackage{transparent}
\usepackage{adjustbox}
\usepackage{flafter}
\usepackage{lscape}
\usepackage{caption}
\usepackage{subcaption}
\usepackage{graphicx}
\usepackage{geometry}
\usepackage{footnote} 
\makesavenoteenv{tabular} 
\usepackage{verbatim}
\usepackage{natbib}
\usepackage{comment}
\usepackage{rotating}
\usepackage{hyperref}
\usepackage{mdframed}
\usepackage{lipsum}
\usepackage{array}
\newcolumntype{H}{>{\setbox0=\hbox\bgroup}c<{\egroup}@{}}
\usepackage{xcolor}
\usepackage{amsmath}
\usepackage{multirow}
\usepackage{bbm}
\usepackage{algorithmicx}
\usepackage{algorithm,algpseudocode}
\usepackage{longtable}
\usepackage{arydshln}

\newcommand{\blind}{0}

\begin{document}

\def\spacingset#1{\renewcommand{\baselinestretch}%
{#1}\small\normalsize} \spacingset{1}

\def\spacingset#1{\renewcommand{\baselinestretch}%
{#1}\small\normalsize} \spacingset{1}


\if0\blind
{
  \title{\bf A Roof Over Risk: A House Price-at-Risk Framework for Hungary\thanks{
    The authors thank Zsuzsanna Hossz\'u, J\'anos Szak\'acs, S\'andor Winkler, Csaba Lados, and Zsuzsanna Husz\'ar for their helpful comments regarding the model, data, and manuscript. We also thank all participants of the RSS 2025, IWAP2025, and MKE 2025 conference. The usual disclaimers apply.}}
  \author{ 
    Tibor Szendrei\\
    National Institute of Economic and Social Research, United Kingdom\\
    \\
    Nikolett V\'ag\'o\\
    Central Bank of Hungary, Hungary\\
     \\
    Katalin Varga\footnote{Corresponding author: vargaka@mnb.hu.}\\
    Central Bank of Hungary, Hungary}
  \maketitle
} \fi

\if1\blind
{
  \bigskip
  \bigskip
  \bigskip
  \begin{center}
    {\LARGE\bf Title}
\end{center}
  \medskip
} \fi


\bigskip
\begin{abstract}
\noindent This paper develops a House Price-at-Risk framework to examine how housing subsidies, credit conditions, and supply factors influence the distribution of house price growth in Hungary. Using quantile regression with adaptive LASSO variable selection, we identify variables driving downside versus upside risks across multiple horizons. Financial stress dominates the lower tail at short horizons, while unemployment and affordability constraints become the primary drivers of downside risk at longer horizons. Housing subsidies exhibit pro-cyclical characteristics, concentrating significant positive effects on the upper quantiles while leaving the lower tail largely unaffected. Supply-side variables display horizon-dependent sign reversals, with construction permits exerting upward pressure on prices in the short run but moderating them as supply materialises. Uncertainty decomposition reveals persistent left-tail dominance across all horizons. These findings suggest that macroprudential frameworks should account for the distributional effects of housing subsidies, particularly their pro-cyclical influence on house price growth.
\end{abstract}


\noindent%
\noindent
\vfill

\spacingset{1.45} 


\section{Introduction}
\input{Chapters/Introductionv2.tex}
\section{House Prices and financial stability}
\input{Chapters/Motiv.tex}

\section{Data} \label{sec:data}
\input{Chapters/Data}

\section{Methodology}
\input{Chapters/Methodology.tex}

\section{House Price-at-Risk models}
\input{Chapters/Resultsv4.tex}

\section{HaR and Financial Stability}
\input{Chapters/FinStab.tex}

\section{Implications for policy}
\input{Chapters/Policy}

\section{Conclusion}
\input{Chapters/Conclusion.tex}

\pagebreak

\bibliographystyle{chicago}
\bibliography{reference.bib}

\pagebreak



\end{document}

%% file: Chapters/Introductionv2.tex
Mortgage and housing loan subsidies represent a ubiquitous policy instrument in contemporary housing markets. Governments employ these mechanisms to address affordability challenges and achieve broader social objectives, particularly around homeownership and family formation. However, the effects of such policies extend well beyond their stated objectives. Housing subsidies can influence not only homeownership rates but also the dynamics of housing markets themselves, with implications for financial stability and macroeconomic outcomes.

Existing research on housing subsidies documents a consistent pattern: house prices frequently increase following the introduction of subsidy programs. Studies from diverse institutional and geographical contexts find that mortgage interest deductions, subsidized loan rates, and family housing allowances systematically elevate housing prices. In the United States, \citet{rappoport2016mortgage} find that eliminating mortgage interest deductions would reduce house prices by 6.9 percent. \citet{gruber2021people} examine Denmark's mortgage interest deduction and find that while it increases housing demand at the intensive margin, leading homeowners to purchase larger and more expensive homes, it has no significant effect on homeownership rates. \citet{vangeel2020influence} document that mortgage interest and capital deduction policies in Belgium increase prices across multiple housing types and regions. \citet{kunovac2022effect} show that Croatian housing loan subsidies increase prices substantially, particularly in developed municipalities. 

This consistent pattern raises an important puzzle. If subsidies succeed in easing borrowing constraints and improving affordability, why do house prices rise in tandem? The conventional explanation emphasizes demand-side effects: subsidies increase the willingness of households to participate in housing markets, but if housing supply responds sluggishly, equilibrium prices adjust upward, potentially eroding much of the affordability benefit. Linneman and Wachter (1989) caution that while mortgage market innovations ease borrowing constraints, policymakers must balance improved affordability against the risk of inflating demand without corresponding supply increases, which can lead to affordability challenges and market volatility.

Existing analyses of housing policy effects focus almost exclusively on conditional mean effects, producing a partial picture of how subsidies reshape housing market dynamics. This conventional approach overlooks a crucial dimension: subsidies may influence different parts of the housing price distribution asymmetrically. It is entirely plausible, for instance, that housing subsidies fuel upside risk of price growth while leaving the bottom tail largely unaffected. In such situations, the mean effects might capture the distribution becoming more skewed, rather than the location of the distribution shifting. Understanding these distributional effects is essential for macroprudential authorities tasked with maintaining financial stability, since housing booms concentrate risk and can generate the leverage dynamics that precede financial crises.

This paper addresses this gap by developing a House Price-at-Risk (HaR) framework tailored to the Hungarian housing market. The HaR approach applies quantile regression methods to characterize the conditional distribution of house price growth across multiple forecast horizons. Rather than estimating a single regression coefficient that averages effects across the distribution, quantile regression permits the estimation of how economic fundamentals influence different quantiles of the conditional price distribution. This richer characterization allows us to distinguish between variables that drive upside risks (rapid price growth) and those that generate downside risks (price declines), and to assess how housing subsidies influence these risks distinctly.

To operationalize this approach, we employ the adaptive LASSO methodology of \citet{szendrei2023revisiting}, which combines quantile regression with variable selection through non-crossing constraints. This technique automatically identifies which variables matter at different points of the distribution, accommodating the possibility of quantile-specific sparsity—that is, economic factors may influence growth prospects only within specific portions of the price distribution.

Hungary provides a compelling case study for this analysis. The country has experienced multiple distinct regimes in housing finance and policy over the past three decades, from state socialist allocation through market liberalization to its recent system of extensive family-oriented subsidies. Moreover, the scale of current interventions is substantial: housing subsidies accounted for approximately 0.6 per cent of GDP in 2023, and the Home Purchase Subsidy (HPS) Scheme alone had supported more than 230,000 families by the end of 2023. In addition, nearly 110,000 families benefited from the interest subsidy associated with HPS loans, while over 240,000 families made use of the Prenatal Baby Support Loan, making the overall impact of these subsidies measurable at the macroeconomic level. Since 2015 Hungarian house prices appreciated 243 percent until the fourth quarter of 2024, when annual growth rate was close to 15 percent creating urgent questions about policy sustainability and financial stability implications.

This paper makes several contributions to the literature. First, it extends the House Price-at-Risk framework to explicitly analyse housing policy effects, filling a gap in the literature on how subsidies reshape the full distribution of price dynamics rather than merely their mean. Second, it provides detailed empirical evidence on how credit and supply-side factors influence different parts of the housing price distribution in the Hungarian context, offering insights relevant to other European economies implementing similar policies. Third, it develops practical tools for macroprudential authorities, demonstrating how HaR measures can be incorporated into early warning systems and dynamic regulatory frameworks such as loan-to-value requirements. The analysis reveals that housing subsidies exert disproportionately large effects on upper quantiles of the price distribution, generating what we term "pro-cyclical" distributional effects that may amplify financial stability risks even as they achieve affordability objectives in the short term.

The remainder of the paper proceeds as follows. Section 2 reviews existing literature on housing supply, financial stability, and the effects of housing subsidies, with particular emphasis on the Hungarian context. Section 3 describes our data and variable construction. Section 4 presents the quantile regression methodology and variable selection approach. Section 5 reports coefficient estimates and out-of-sample forecasting performance across multiple model specifications and forecast horizons. Section 6 develops measures of uncertainty and skewness decomposition over time and analyses the spillovers between housing market tail risks and systemic financial risk. Section 7 concludes with implications for macroprudential policy.

%% file: Chapters/Motiv.tex
\subsection{Impact of housing supply on house price dynamics}

The impact of housing supply on house price dynamics has been of interest in the literature. In the paper by \citet{been2024supply}, the authors argue that a key challenge when it comes to studying the impacts of housing supply on rents and prices, is isolating the effects new housing has, because property development tends to occur at locations that are already poised for growth. To this end there is a selection bias, which leads to observed rent changes potentially reflecting pre-existing demand rather than the direct influence of additional supply. A key contribution of this paper is that it emphasises empirical strategies to address the endogeneity inherent in modelling housing supply and demand.

\citet{ooi2013spillover} highlights that the net effect of new infill developments hinges on the balance of two channels: 1) the `amenity effect', and; 2) `supply effect'. The first effect is an increase in demand as modern designs and improved aesthetics boost the demand for the neighbourhood, which in turn attracts higher-income residents. The second effect, is the traditional increase in supply effect, i.e. new housing units increases the inventory which leads to increased competition among properties and thus lower prices. The overall impact of new housing thus depends on which of these opposing forces prevails.

\citet{gonzalez2022spillover} studies the localised spillover effects of new housing on property prices. The authors find that while an increase in the housing supply would, ceteris paribus, put downward pressure on prices due to the added competition, there are several demand factors that work against this. Importantly, in some cases the demand factors can reverse the effect of the increased housing supply. These reversal in effects can happen, on account of displacement, i.e. the case when high-quality housing attracts wealthier housholds. This in turn can lead to a demand for better local amenities (i.e. the `amenity effect' of \citet{ooi2013spillover}) which in turn stimulates demand. In the case study from Montevideo, Uruguay, the dynamics mentioned led to a 12\% increase in surrounding house prices. Importantly, this increase in prices were observed within a 200 meter radius.

\subsection{Upside risk of house prices on financial stability}

The interplay of the real estate market and financial stability has been a topic of interest ever since the global financial crisis of 2008. \citet{kuttner2016can} highlights that the sole reliance on interest rate policy for housing market stabilization is insufficient to achieve financial stability goals without negative impacts on the macroeconomy. Due to the interest rate tool being `blunt', the authors advocate for the use of more targeted macroprudential policies to achieve financial stability. As such, it is not surprising that most papers that look at the interaction between the real estate and financial markets highlight the role macroprudential policy can have in moderating real estate booms and mitigating systemic risk.

\citet{ely2021transmission} demonstrate that borrower-based policy tools, such as limits on the loan-to-value (LTV) or debt-to-income (DTI), can `lean' against real estate booms. This in turn can curb excessive house price dynamics which in turn limits volatility in the macroeconomy. This is also the key argument of \citet{hartmann2015real}, who notes that the cyclical nature of real estate markets has been a large contributor to systemic financial crises in the past. This finding has been further corroborated by \citet{jorda2015leveraged}, who find that credit market booms coupled with asset price booms (such as housing) leads to deeper and longer lasting recessions.

\citet{zhang2016leaning} looks at how Asian economies, which have introduced macroprudential policies after the east Asian crisis of 1996-1997, fared in maintaining financial stability. Their key finding is that certain macroprudential instruments have been successful in mitigating the accummulation of financial risks during periods of economic growth. In particular, housing-related macroprudential instruments effectively reduced credit growth, curbed housing price inflation, and limited bank leverage. Complementing these findings, \citet{morgan2019ltv} discusses the broader utility of macroprudential policies. In particular they highlight that not only do LTV and DSTI ratios limit credit growth, but they also limit banks' exposure to high risk loans, that are likely to turn non-performing during economic downturns. The authors point out that while many studies have focused on the impact macroprudential policies have on credit growth, the application of many macroprudential policy levers leads to lower financial stability risks through reducing risk on banks' balance sheets.

\subsection{Impact of subsidies on house price dynamics}

\citet{linneman1989impacts} shows that mortgage market innovations have helped in easing borrowing constraints and improving house affordability. As such, mortgage subsidies have the potential to enhance housing affordability by easing borrowing constraints. However, the authors caution that the increased affordability must be balanced against the risk of inflating demand without corresponding increases in supply, which can lead to affordability challenges and market volatility. Due to this trade-off, there have been several papers that study the impact of mortgage subsidies on housing market dynamics.

\citet{greenwald2021credit} look at the interaction between credit conditions and house prices in the US. The paper uses three different credit supply instruments to estimates the relative elasticities of the price-rent ratio and homeownership with respect to credit shocks. The findings suggest that rental markets are highly frictional and closer to fully segmented, implying significant effects of credit on house prices. Changes in credit standards can explain between 34\% and 55\% of the rise in price-rent ratios during the housing boom.

This dual impact of mortgage and housing loan subsidies on housing markets has been analysed by \citet{rappoport2016mortgage}. The paper highlights that subsidies benefit buyers by reducing mortgage interest payments but can also increase house prices. The net effect on buyers depends on the loan-to-value (LTV) ratio and the elasticity of housing supply and demand. In an empirical application of the US metropolitan areas, the authors finds that eliminating mortgage interest deductions would reduce house prices by 6.9\%.

The United States is not the only country where the interplay between credit market and housing market has been studied. \citet{gruber2021people} examines the long-term effects of morgage subsidies on housing decisions using data from Denmark. They have various findings: 1) mortgage interest deduction has no significant effect on the rate of homeownership, even in the long run; 2) deduction significantly increases housing demand at the intensive margin, leading homeowners to buy larger and more expensive houses; and 3) largest impact of the mortgage interest deduction is on household financial decisions, particularly increasing indebtedness. The findings suggest that while the mortgage interest deduction distorts housing and debt demand, it is ineffective at promoting homeownership and any associated externalities.

\citet{kunovac2022effect} studies the impact of housing loan subsidies in Croatia. They find that after the introduction of the subsidy for which households could apply only during a one-month period, house prices increased especially in developed municipalities. Furthermore, the authors find no effect on the aggregate number of transactions and on homeownership rates but find a disruption of the usual intra-annual dynamics of housing transactions, as housing transactions concentrated around in specific months.

\citet{akgunduz2023cost} looks at the impact of reduced mortgage rates on house prices in Turkey, using a subsidized mortgage program as a case study. The authors find that a 1 percentage point decrease in mortgage rates led to a 3.3\% increase in individual mortgage loans and a 1.6\% increase in house prices. Furthermore, the program resulted in a significant increase in the LTV ratios of mortgage loans. 

\citet{vangeel2020influence} look at how the mortgage interest and capital deduction policy affects house prices in different regions and for various housing types in Belgium. Overall, the policy has been found to increase prices for housing. The study also differentiates between various types of housing and find heterogeneity in the impact: prices for ordinary houses and apartments increased while the price for villas has decreased. The authors also find that the impact of the policy varies across the different regions that were studied, highlighting that local housing markets can have distinct responses to fiscal incentives.

These studies collectively highlight the multifaceted effects of mortgage and housing loan subsidies on housing markets. While subsidies can improve affordability, they often lead to increased house prices.

\subsection{Housing Loans in Hungary}

Hungary's housing finance system has undergone a remarkable transformation since 1989, creating an environment particularly well-suited for House Price-at-Risk analysis. The country's evolution from state socialism through market liberalization to extensive subsidy-based intervention provides a rich laboratory for understanding how housing policies influence price dynamics and financial stability.

Prior to 1990, the state-owned OTP Bank maintained a monopoly on housing finance, controlling approximately 85\% of the loan portfolio. The transition to a market-based system created what \citet{hegedus2000crisis} have described as a fundamental crisis in housing finance that persisted throughout the 1990s. The housing privatization program of the early 1990s established the foundation for Hungary's current homeownership structure, with approximately 35\% of public housing stock privatized by 1993 at prices between 15-40\% of market value. Following World Bank recommendations, Hungary began reconstructing mortgage markets in 1996, supported by legal framework improvements that reduced enforcement times from years to merely 30 days \citep{czirfus2022, hegedus2000crisis}.

In the late 1990s high interest rates and limited bank lending kept household indebtedness low \citep{hegedus&teller2005}. Home-ownership remained high due to the post-socialist legacy, but access to new housing finance was restricted and uneven, favouring higher-income households \citep{hegedus2007}. A significant shift occurred in 2000, when the government introduced a generous, state-subsidized mortgage programme that greatly lowered borrowing costs and broadened eligibility \citep{hegedus&somogyi2016}. After 2002, financial sector liberalization and the entry of foreign banks intensified mortgage competition and increased credit availability, including to lower-income households \citep{hegedus&somogyi2016}. During the credit boom state-subsidized mortgages and the expansion of foreign-currency lending disproportionately fuelled a surge in home purchases, construction, and household indebtedness in economically stronger urban regions, with Budapest and other major cities capturing higher benefits \citet{posfai2018crisis}. Although the subsidy scheme was curtailed in 2004 due to fiscal pressures by then the mortgage market had already expanded significantly, establishing a debt-based model of home-ownership and laying the structural foundations for subsequent financial vulnerabilities \citep{bohle2013}.

The period from 2004 to 2015 represents perhaps the most dramatic episode in Hungarian housing finance history. The introduction of Swiss franc mortgages in 2004, coinciding with EU accession, created significant vulnerabilities in the system. These products offered approximately 1.5 percentage points lower interest rates than euro-denominated loans, making them attractive despite imposing extreme exchange rate risk on borrowers \citep{gagyi2023fx}. The 2008-2009 financial crisis exposed these vulnerabilities dramatically, as the forint depreciated 26\% against the euro and 66\% against the Swiss franc, causing CHF loan instalments to increase by 70-80\%. Unlike other Central European countries where foreign exchange lending primarily affected upper-middle classes, Hungary showed deeper social penetration, affecting households taking loans for basic housing needs rather than investment purposes.

The parliamentary ban on new foreign exchange mortgage lending in July 2010 marked the beginning of systematic crisis management, followed by early repayment schemes in 2011-2012 that allowed 170,000 loan holders to repay debts at predetermined favourable rates. The crisis revealed a fundamental vulnerability in the Hungarian housing market: many households seeking housing were driven to high-risk financial products due to a lack of affordable alternatives. The tighter post-crisis credit regulation can increase financial stability but has the unintended potential to constrain market-based access to housing finance. Against this backdrop, the Hungarian authorities increasingly turned to extensive family-oriented housing subsidies as the primary mechanism for supporting home-ownership.

The Home Purchase Subsidy Scheme for Families (HPS), introduced in 2015, represents Hungary's flagship housing policy and one of Europe's most generous subsidy systems. Since inception, the program has allocated 762 billion HUF to 232,000 families until the end of 2023. The program's structure explicitly links housing support to demographic objectives. The largest benefit within the program is available for married couples  with three or more children receiving maximum benefits of 10 million HUF in non-refundable grants for new homes, plus subsidized loans up to 15 million HUF at 3\% interest rates compared to market rates of 6-9\%. Additional programs, including the Prenatal Baby Support Loan providing up to 10 million HUF in general-purpose loans that become interest-free after the first child and fully forgiven after the third child, reinforce this family-focused approach \citep{sagi2022key}. By 2023, housing subsidies reached 0.6\% of GDP, representing an extraordinary level of government intervention in the housing market.\footnote{For further details please see \url{https://www.mnb.hu/en/publications/reports/public-finance-report}}

The combined effects of crisis resolution and subsidy programs have produced remarkable price growth. Hungarian housing prices have increased 243\% between 2015 Q4 and 2024 Q4, with recent performance showing 14.7\% year-on-year growth nationwide in the fourth quarter of 2024, reaching 14.1\% in Budapest. This dramatic price appreciation raises important questions about the sustainability of current housing system and implications for affordability and financial stability. Also note that housing policy outcomes vary significantly across regions of Hungary, with Budapest and surrounding areas capturing disproportionate benefits.

Hungary provides exceptional value for House Price-at-Risk research due to several distinctive characteristics. First, the country offers comprehensive policy experimentation that allows researchers to study subsidy impacts across multiple instruments and time periods. From market-based systems to extensive family subsidies, Hungary provides natural experiments in policy effectiveness that are rare in developed economies. Second, data availability and institutional transparency enable rigorous empirical analysis. Hungary maintains comprehensive housing statistics and the Hungarian Central bank regularly analyse the Housing market developments facilitating longitudinal studies of policy impacts that would be difficult to replicate elsewhere. Third, the scale and scope of interventions make effects measurable at the macroeconomic level. With subsidies representing significant percentages of GDP during peak periods, impacts are observable in aggregate statistics rather than limited to micro-level analysis.

The Hungarian experience demonstrates how housing policy can serve multiple objectives beyond shelter provision, while simultaneously illustrating the complex feedback mechanisms between policy interventions, price dynamics, and financial stability. The country's trajectory from foreign exchange mortgage crisis through extensive subsidy implementation provides a particularly valuable case study for understanding how demand-side interventions influence the conditional distribution of house price growth, making it an ideal setting for developing and testing House Price-at-Risk models that can inform macroprudential policy.

%% file: Chapters/Data.tex
The empirical analysis incorporates both demand-side and supply-side fundamentals that influence residential real estate prices in Hungary. The dataset includes the Factor-based Index of Systemic Stress (FISS), developed by \citet{szendrei2020fiss}, which captures financial market conditions that may exert short-term pressure on house prices. The non-stationary factor methodology has been shown to yield good performance in capturing financial stress in the UK too \citep{varga2024non}.

\begin{table}[t!]
\caption{Table of variables}
\label{tab:varibs}
\resizebox{\textwidth}{!}{%
\begin{tabular}{l|c|l}
\hline
\multicolumn{1}{c|}{\textbf{Short name}} & \textbf{Model version}     & \multicolumn{1}{c}{\textbf{Details}} 
\\ \hline
\textbf{HCPI}                             & HCPI QoQ                   & Real MNB House Price Index \\
\textbf{FISS}                             & FISS                       & Factor based Index of Systemic Stress \\
\textbf{New credit plus Benefits}         & New credit plus Benefits   & Newly issued mortgage loans plus Home Purchase Subsidy (HPS) \\ & &Scheme for Families, rural HPS, HPS loans, rural HPS loans and \\ & &Prenatal Baby Support Loans\\
\textbf{Unemployment rate}                & UnempRate (YoY)            & Long-term unemployment rate\\
\textbf{Income}                           & Income (YoY)               & Real personal disposable income \\
\textbf{Price-to-Income Gap}              & Price-to-Income Gap        & Deviation of price-to-income from its long-term average \\
\textbf{Building costs}                   & Building Costs (YoY/QoQ)   & Construction cost index (2015=100\%) \\
\textbf{Construction permits}             & Construction Permits (YoY) & Number of building permits \\
\hline
\end{tabular}%
}

\end{table}

Demand-side fundamentals encompass macroeconomic indicators such as household income and unemployment rates, alongside variables capturing credit market conditions. The Price-to-Income Gap measures housing affordability relative to its long-term average, providing insight into valuation pressures. Credit channel effects are captured through newly issued mortgage volumes and housing subsidy programmes, which influence both investment decisions and credit availability for prospective homebuyers.

Supply-side fundamentals include construction costs, measured through a construction cost index based on 2015 prices, and the number of building permits issued, which serves as a forward-looking indicator of housing supply. These variables reflect capacity constraints, input cost pressures, and regulatory factors affecting the construction sector's ability to respond to demand shifts.

The dataset spans from 2005 Q1 to 2024 Q4 at quarterly frequency, ensuring a balanced panel that captures multiple phases of the Hungarian housing cycle. All variables underwent stationarity testing, with integrated variables of order one transformed to achieve stationarity. For persistent variables, we evaluated quarter-on-quarter, year-on-year, and biannual growth rates. Variables exhibiting seasonal patterns were transformed using year-on-year differences to eliminate seasonal components. Table 1 presents the final set of regressors alongside their transformations.

To isolate the effects of credit and subsidy policies, we construct three specifications of the credit variable. The first specification includes only newly issued mortgage loans. The second incorporates the Home Purchase Subsidy (HPS) Scheme for Families, rural HPS alongside standard mortgages  and (rural) HPS loans. The third specification adds the Prenatal Baby Support Loan programme, which requires no repayment upon the birth of at least one child, with the benefit amount varying according to the number of children born. This tiered approach allows us to assess whether housing subsidies amplify the relationship between credit conditions and house price dynamics beyond the effect of conventional mortgage lending alone.

%% file: Chapters/Methodology.tex
\subsection{Non-Crossing Quantile Regression}

Quantile regression, introduced by \citet{koenker1978regression}, extends least absolute deviation regression by providing lines of best fit for different quantiles of a response variable ($Y$) distribution. Similar to traditional regression, the objective is to estimate conditional quantiles based on a vector of predictors $X \in \mathbb{R}^K$. Mathematically, the $q^{th}$ quantile is expressed as:

\begin{equation} \nonumber
    \mathcal{Q}_q=X^T\beta_{\tau_q}.
\end{equation}

The parameters for the $Q$ quantiles ($\beta = \{\beta_{\tau_1}, \beta_{\tau_2}, \ldots, \beta_{\tau_Q}\}$) characterize the conditional distribution. The goal of quantile regression is to estimate the coefficient vector $\beta_{\tau_q} \in \mathbb{R}^K$ for all quantiles. This is achieved by solving the quantile regression minimization problem:

\begin{equation}\label{eq:QR}
    \hat{\beta}=\underset{\beta}{argmin}\sum^{Q}_{q=1}\sum^{T}_{t=1}\rho_q(y_t-x_t^T\beta_{\tau_q})
\end{equation}

\noindent where the tick-loss function $\rho_q(u)$ is defined as:

\begin{equation}
    \rho_q(u)=u(p-I(u<0))
\end{equation}

This asymmetric loss function assigns different weights to positive and negative residuals, ensuring that $\tau_q$ proportion of the observations fall below and $(1-\tau_q)$ fall above the quantile regression line.

A challenge in quantile regression arises from quantile crossing, i.e. a situation where estimated quantiles are not monotonically increasing. This violates the monotonicity assumption and often occurs due to data scarcity, particularly in time-series datasets \citep{koenker2005}. To address this, two main approaches have been proposed: (1) fitting a distribution for each time period individually based on the estimated quantiles \citep{adrian2019vulnerable, korobilis2017quantile}, or (2) sorting the quantiles post-estimation to ensure monotonicity \citep{chernozhukov2009improving}. While these methods correct the fitted quantiles, they do not adjust the underlying coefficients, which lead \citet{bondell2010noncrossing} to propose an estimator that ensures non-crossing quantiles in-sample.

Non-crossing constraints can be incorporated directly into the quantile regression framework to enforce monotonicity of the estimated quantiles. This is achieved by introducing inequality constraints into the optimization problem:

\begin{equation} \label{eq:NCQR}
    \begin{split}
        \hat{\beta}&=\underset{\beta}{argmin}\sum^{Q}_{q=1}\sum^{T}_{t=1}\rho_q(y_{t}-x_t^T\beta_{\tau_q})\\
        &s.t.~x^T\beta_{\tau_q} \geq x^T\beta_{\tau_{q-1}}
    \end{split}
\end{equation}

While conceptually straightforward, this approach introduces $T \times (Q - 1)$ inequality constraints, which can complicate estimation. To address this, \citet{bondell2010noncrossing} propose restricting the domain of interest to $\mathcal{D} \in [0,1]^K$ and focusing on the worst-case scenario in the data.\footnote{This refers to the situation where the negative difference coefficient's ($\gamma^-_{j,\tau_q}$) corresponding variable values are 1, and the positive difference coefficients ($\gamma^+_{j,\tau_q}$) corresponding variables equal 0.} This reduces the number of constraints to $(Q - 1)$.

The coefficients are then reparameterized in terms of quantile differences: $(\gamma_{0,\tau_1}, \ldots, \gamma_{K,\tau_1})^T = \beta_{\tau_1}$ for $q=1$, and, $(\gamma_{0,\tau_q}, \ldots, \gamma_{K,\tau_q})^T = \beta_{\tau_q} - \beta_{\tau_{q-1}}$ for $q = 2, \ldots, Q$. Under this reparameterization, the non-crossing constraint becomes:

\begin{equation}
    x^T\gamma_{\tau_q} \geq 0
\end{equation}

Further, decomposing the $j^{th}$ quantile difference as $\gamma_{j,\tau_q} = \gamma^+_{j,\tau_q} - \gamma^-_{j,\tau_q}$, where $\gamma^+_{j,\tau_q}$ and $-\gamma^-_{j,\tau_q}$ are its positive and negative parts, respectively, leads to a simplified constraint:

\begin{equation}
    \gamma_{0,\tau_{q}}\geq \sum^K_{j=1}\gamma^-_{j,\tau_q} ~ (q=2,...,Q) \label{eq:constraint}
\end{equation}

This ensures that the sum of negative shifts does not reduce the quantile below the change in intercept, which serves as a location shifter. \citet{bondell2010noncrossing} demonstrate that this condition is both necessary and sufficient for non-crossing quantiles. 

\citet{szendrei2023fused} highlight the equivalence between non-crossing constraints and interquantile shrinkage. As such these constraints, beyond enforcing quantile monotonicity, are helpful in identifying quantile varying and quantile constant covariates when jointly estimating the quantiles. However, joint estimation struggles in the presence of quantile specific sparsity, a situation where a covariate has an impact only for a (potentially very limited) subset of the quantiles being estimated \citep{kohns2021decoupling,kohns2025joint}. Quantile specific sparsity is a special case of quantile varying coefficient: rather than a continuous coefficient profile, the quantile specific sparse coefficient profile is discontinuous. To this end we will impose shrinkage on the levels of the coefficients to tackle cases where quantile specific sparsity is present. 

\subsection{LASSO}
There are several variable selection methods available that can be used to shrink the levels of the coefficients. In this paper we will opt to use the adaptive LASSO. The choice was motivated for two reasons. First, we wish to sparsify the design matrix so as to identify that could be quantile specific sparse. This will allow us to identify which variables have an influence on the different parts of the distribution. 
Second, the LASSO shrinkage in quantile regression has been studied and applied extensively (see: \citet{belloni2011ell}, \citet{jiang2013interquantile}, \citet{jiang2014interquantile}, and \citet{szendrei2023revisiting} among others).

In this paper we will use the quantile LASSO estimator of \citet{szendrei2023revisiting}. In this formulation, variable selection is imposed `globally' (there is one overall tuning parameter) along with non-crossing constraints. Using only one tuning parameter helps with computational feasibility, while imposing non-crossing constraints ensures that the LASSO induced quantile specific sparsity \citep{kohns2021decoupling} does not lead to our estimated quantile curves crossing. The final estimator is the following:

\begin{equation} \label{eq:LASSONC}
    \begin{split}
        \hat{\beta}(\tau)=&\underset{\beta,\alpha}{min}\sum^Q_{q=1}\sum^n_{i=1}\rho_{\tau_q}(y_i-x_i^T\beta_{\tau_q})\\
        &s.t.~ \gamma_{0,\tau_{q}}\geq \sum^K_{j=1}\gamma^-_{j,\tau_q} \\
        &\sum^Q_{q=1}\sum^K_{k=2}w_{k,\tau_q}|\beta_{k,{\tau_q}}|\leq t^*
    \end{split}
\end{equation}

\noindent where we set the weights following \citet{jiang2014interquantile}: $w_{k,\tau_q}=|\theta_{k,\tau_q}|^{-1}$, where $\theta$ are the estimated coefficients of a regular quantile regression using the full design matrix. The (adaptive) LASSO constraint regulates the amount of variation the coefficients can have in each quantile by constraining the sum of the coefficients to be at most $t^*$.

There are several ways one can select the optimal $t^*$. Cross-Validation is a popular choice for obtaining optimal tuning parameters. The key idea of Cross-Validation is to break down the sample into a training sample (where coefficients are estimated) and a testing sample (where model fits are evaluated). This way one is able to achieve a model that does not overfit. One potential drawback of Cross-Validation, is that it requires the sample to be broken down. To this end the results of \citet{stone1977consistent} and \citet{shao1997asymptotic} are critical. The findings of these papers show that one can rely on information criteria for hyperparameter selection. As such, given the limited number of observations we opt to use information criteria to obtain the optimal $t^{*}$.
These methods look at in-sample fit only but penalise for model complexity:

\begin{equation} \label{eq:IC}
    IC(t^*)=\sum^Q_{q=1}log\Big[\sum^n_{i=1}\rho_{\tau_q}(y_i-\alpha_{\tau_q}(t^*)-x_i^T\beta_{\tau_q}(t^*))\Big]+P(t^*)
\end{equation}

\noindent where $P(t^*)$ is some penalty function penalising model complexity. The penalty function for the AIC and BIC are taken from \citet{jiang2014interquantile}: 

\begin{equation} \label{eq:AICBIC}
    \begin{split}
        P_{BIC}(t^*)&=\frac{log(n)}{2n}\sum^Q_{q=1} \sum^K_{j=1} \mathbb{I}(\beta_{\tau_q,j}(t^*)>0)
    \end{split}
\end{equation}

When substituting equations (\ref{eq:AICBIC}) into equation (\ref{eq:IC}) it is clear that there is a trade-off between model fit and the number of parameters included. Since the coefficients (and penalty) are a function of $t^*$ from equation (\ref{eq:LASSONC}), by running several candidate values of $t$, we are able to obtain the penalised fit for the different values of $t$. By looking for the value of $t$ that minimises this loss function, we can find the optimal number of parameters that explain the variation in the data. 

We follow \citet{szendrei2023revisiting} and impose a global hyperparameter rather than a quantile specific one. On account of this we can use grid search to obtain the optimal $t^*$ parameter. A benefit of grid search is that it is embarrassingly parallel which speeds up computation. For each candidate hyperparameter, equation (\ref{eq:LASSONC}) is run and their respective AIC and BIC are recorded. The model with the lowest information criteria (BIC) is chosen as the optimal model.

When doing regularised regression it is common to standardise the data, so as to make the scales equal for each variable. However, to impose non-crossing constraints we already used the minimax transformation to make our variables be on the domain of [0,1] \citep{bondell2010noncrossing}. As such no further data transformation is necessary.


%% file: Chapters/Resultsv4.tex
\begin{table}[h!]
\caption{Overview of models}
\label{tab:ModOverview}
\centering
\begin{tabular}{l|ccc}
\hline
 &  Full & Supply & Credit \\ \hline
HP lag & x & x  & x \\
FISS &  x & x  & x \\
Price/Income Gap & x & x  & x \\
Income  & x & x  & x\\
Unemployment Rate  & x & x  & x \\ \hdashline
New credit plus Benefits & x & & x \\ \hdashline
Construction permits  & x & x &  \\ 
Building costs  &  & x & \\ \hline
\end{tabular}%

\end{table}

The model is run on the variables described in section (\ref{sec:data}). The variables were chosen to have a wide range of effects. The various effects the variables capture can occur at different horizons. To this end we will estimate the model over various horizons, namely $h=1Q$, $h=4Q$, (i.e. 1 year ahead), and $h=8Q$ (i.e. 2 years ahead). We estimate the model described in equation (\ref{eq:LASSONC}) for every $10^{th}$ quantile. Beyond looking at the coefficients we will also look at in sample fits, and a small forecast exercise is conducted to evaluate out-of-sample performance of the different models. 

Two further versions of the model are estimated, 1 without credit variable (i.e. a housing supply focused model) and 1 without housing supply variables (i.e. a housing credit focused model). By estimating 3 models, we get a better idea of how much the model improves its out-of-sample performance as the different segments are integrated in the estimation. Table (\ref{tab:ModOverview}) summarises the different models and the variables that are included.

\subsection{Coefficients}

The coefficients of these models for various quantiles are presented in table (\ref{tab:ModelRuns}), along with the OLS coefficients for the models as well. Throughout this table, zeros indicate the variable was not selected by LASSO.

House price lag is left unshrunk across quantiles and horizons following the literature on other macroeconomic at-risk models (e.g. \citet{adrian2019vulnerable} and \citet{szendrei2023revisiting}). It is always positive, but shows little cross-quantile variation at a given horizon; the main variation is across horizons, where the lag coefficient declines markedly (Full: 0.44 at $h=1$ to 0.145 at $h=8$).House price lag is important but shows little quantile variation at a given horizon. This is not the case across forecast horizons, where the magnitude of the lag coefficient drops substantially. This weaker dependence on past prices at longer horizons is likely on account of other macroeconomic and supply-side factors being influential for growth dynamics at these horizons. The OLS estimates provide a useful baseline and largely confirm the patterns across the forecast horizon. 

The variable capturing financial stress (FISS) shows consistently negative coefficients, with very large short run effects. In the Full model, FISS at $h=1$ is -2.20 at $\tau=0.1$ and remains negative across the other quantiles (e.g. -1.86 at $\tau=0.3$), implying substantial short-run downside risk that diminishes across quantiles and forecast horizon. This indicates that financial stress disproportionately suppresses price growth of the housing market. This is likely on account of heightened sensitivity among financially constrained or risk-averse buyers. The FISS coefficient at $h=1$ for $\tau=0.1$ is large relative to the other variable's coefficients. This implies very large short-run downside risk is associated with financial stress, which weakens as we move up the quantiles (and further in the horizon). The reduced importance of FISS at higher quantiles ($\tau=0.7$ and $\tau=0.9$) for longer horizons suggests that instances of rapid price growth are less sensitive to financial stress. Financial stress being less important for the long term horizon is in line with the findings of \citet{varga2024non}, namely that factor based measures of financial stress are better suited for capturing risks that materialise in the short run. The OLS estimates corroborate the quantile findings by revealing a negative relationship between financial stress and house price growth.

\begin{landscape}
\begin{table}[]
\caption{QR and OLS coefficients of the models}
\label{tab:ModelRuns}
\centering
\resizebox{1.3\textwidth}{!}{%
\begin{tabular}{cr|ccc|ccc|ccc|ccc|ccc||ccc}
\hline
 &  & \multicolumn{3}{c|}{$\tau=10$} & \multicolumn{3}{c|}{$\tau=30$} & \multicolumn{3}{c|}{$\tau=50$} & \multicolumn{3}{c|}{$\tau=70$} & \multicolumn{3}{c||}{$\tau=90$} & \multicolumn{3}{c}{OLS} \\
 &  & Full & Supply & Credit & Full & Supply & Credit & Full & Supply & Credit & Full & Supply & Credit & Full & Supply & Credit & Full & Supply & Credit \\ \hline
\multicolumn{2}{l|}{h=1Q} &  &  &  &  &  &  &  &  &  &  &  &  &  &  &  &  &  &  \\
 & Intercept & -2.4377 & -2.7489 & -2.3469 & -0.6336 & -0.3785 & -0.6142 & 0.8089 & 1.0317 & 0.9995 & 2.0627 & 1.7993 & 1.775 & 2.8437 & 2.8495 & 3.0812 & 0.5675 & 0.5675 & 0.5675 \\
 & HP lag & 0.4446 & 0.4838 & 0.4813 & 0.4446 & 0.4838 & 0.4813 & 0.4366 & 0.4838 & 0.4671 & 0.4366 & 0.3621 & 0.5354 & 0.4366 & 0.3621 & 0.5354 & 0.4230 & 0.5849 & 0.4677 \\
 & FISS & -2.2012 & -2.0681 & -2.2558 & -1.8602 & -1.8291 & -1.8572 & -1.2309 & -1.5182 & -1.3458 & -1.3351 & -1.4615 & -1.3861 & -1.4816 & -1.3162 & -1.3861 & -1.7122 & -1.7149 & -1.6762 \\
 & Price/Income Gap & -0.7277 & -0.2799 & -0.6471 & 0 & 0 & 0 & 0 & 0 & -0.1508 & 0 & 0 & 0 & 0 & 0 & 0 & -0.3433 & -0.3860 & -0.4141 \\
 & Income & 0 & 0 & 0 & 0 & 0 & 0 & 0 & 0 & 0 & 0 & 0 & 0 & 0 & 0 & 0 & -0.1056 & -0.1657 & 0.0638 \\
 & Unemployment & 0 & 0 & 0 & 0 & 0 & 0 & 0 & 0 & 0 & 0 & 0 & 0 & 0 & -0.2155 & -0.5717 & 0.0961 & 0.1444 & 0.0011 \\ \hdashline
 & New credit plus Benefits & 0 &  & 0 & 0 &  & 0 & 0 &  & 0.5161 & 0.5637 &  & 0.6042 & 0.5357 &  & 0.6042 & 0.2301 &  & 0.2637 \\ \hdashline
 & Construction permit & 0 & 0 &  & 0 & 0.3211 &  & 0.444 & 0.4226 &  & 0.6304 & 0.6052 &  & 0.6463 & 0.5751 &  & 0.4321 & 0.4173 &  \\
 & Building cost &  & 0 &  &  & 0.4259 &  &  & 0.7513 &  &  & 0.8399 &  &  & 0.7257 &  &  & 0.6415 &  \\ \hline
\multicolumn{2}{l|}{h=4Q} &  &  &  &  &  &  &  &  &  &  &  &  &  &  &  &  &  &  \\
 & Intercept & -1.1566 & -1.211 & -1.0719 & -0.3603 & -0.4046 & -0.3948 & 0.7644 & 0.6841 & 0.3644 & 1.6323 & 1.4055 & 1.6538 & 2.2728 & 2.1232 & 2.2695 & 0.5526 & 0.5526 & 0.5526 \\
 & HP lag & 0.4036 & 0.284 & 0.3529 & 0.2265 & 0.284 & 0.2882 & 0.2265 & 0.284 & 0.2276 & 0.2265 & 0.284 & 0.2276 & 0.2265 & 0.284 & 0.2276 & -0.0003 & 0.0488 & 0.0319 \\
 & FISS & -1.6597 & -1.642 & -1.7025 & -1.6597 & -1.642 & -1.7025 & -1.4182 & -1.6104 & -1.3854 & -1.4066 & -1.4511 & -1.3854 & -1.1025 & -1.4511 & -1.0945 & -1.4431 & -1.5039 & -1.4190 \\
 & Price/Income Gap & -0.6537 & -0.7062 & -0.6148 & -0.6537 & -0.7062 & -0.6148 & -0.6537 & -0.7062 & -0.6762 & -0.6537 & -0.7113 & -0.6762 & -0.7308 & -0.7364 & -0.7449 & -0.4965 & -0.5602 & -0.5442 \\
 & Income & 0 & 0 & 0 & 0 & 0 & 0 & 0 & 0 & 0 & 0 & 0 & 0 & 0 & 0 & 0 & -0.0427 & 0.0096 & 0.0636 \\
 & Unemployment & -0.654 & -0.5975 & -0.8175 & -0.4879 & -0.3943 & -0.4738 & -0.4879 & -0.1188 & -0.4125 & 0 & 0 & 0 & 0 & 0 & 0 & -0.4166 & -0.3625 & -0.4723 \\ \hdashline
 & New credit plus Benefits & -0.1077 &  & 0 & 0 &  & 0 & 0.5068 &  & 0 & 0.5633 &  & 0.5751 & 0.6339 &  & 0.6555 & 0.3051 &  & 0.3308 \\ \hdashline
 & Construction permit & 0 & 0 &  & 0.0656 & 0.1843 &  & 0 & 0.1843 &  & 0 & 0.1378 &  & 0 & 0 &  & 0.2773 & 0.2496 &  \\
 & Building cost &  & 0 &  &  & 0.1652 &  &  & 0.5758 &  &  & 0.6603 &  &  & 0.621 &  &  & 0.4073 &  \\ \hline
\multicolumn{2}{l|}{h=8Q} &  &  &  &  &  &  &  &  &  &  &  &  &  &  &  &  &  &  \\
 & Intercept & -1.1013 & -1.1793 & -1.1081 & -0.0383 & 0.1409 & -0.0069 & 0.6829 & 0.7441 & 0.7622 & 1.2639 & 1.1279 & 1.1813 & 2.062 & 2.2098 & 2.0912 & 0.5750 & 0.5750 & 0.5750 \\
 & HP lag & 0.1447 & 0.1451 & 0.1092 & 0.1447 & 0.1451 & 0.1092 & 0.1447 & 0.1451 & 0.0842 & 0.1447 & 0.1451 & 0.0842 & 0.1447 & 0.1451 & -0.169 & 0.1616 & 0.1255 & 0.0751 \\
 & FISS & -0.2197 & -0.2173 & -0.4978 & 0 & 0 & 0 & 0 & 0 & 0 & 0 & 0 & 0 & 0 & 0 & 0 & -0.0260 & -0.0646 & -0.0426 \\
 & Price/Income Gap & -0.9993 & -1.0619 & -1.1674 & -0.9993 & -1.0619 & -0.9812 & -0.9182 & -1.0619 & -0.911 & -1.0064 & -1.0619 & -0.9969 & -1.0414 & -0.9199 & -1.0673 & -1.0687 & -1.1162 & -1.0071 \\
 & Income & 0.9022 & 0.9331 & 0.7515 & 0.9022 & 0.9331 & 0.7515 & 0.8869 & 0.9331 & 0.7807 & 0.8869 & 0.8782 & 0.7899 & 0.8869 & 0.7893 & 0.7899 & 0.8669 & 0.8740 & 0.7385 \\
 & Unemployment & -1.2165 & -1.1419 & -1.2032 & -1.1058 & -1.1419 & -1.1175 & -1.0868 & -0.9568 & -0.9926 & -0.9483 & -0.9787 & -0.8545 & -0.9198 & -0.9787 & -0.6929 & -1.1449 & -1.1210 & -1.0893 \\ \hdashline
 & New credit plus Benefits & 0 &  & 0 & 0 &  & 0 & 0 &  & 0.2991 & 0.2034 &  & 0.3301 & 0.2034 &  & 0.3667 & 0.2159 &  & 0.2203 \\ \hdashline
 & Construction permit & 0 & 0 &  & -0.156 & -0.1247 &  & -0.2875 & -0.2051 &  & -0.2875 & -0.2606 &  & -0.465 & -0.3881 &  & -0.3668 & -0.3535 &  \\
 & Building cost &  & 0 &  &  & 0.3373 &  &  & 0.3373 &  &  & 0.3373 &  &  & 0.4233 &  &  & 0.3208 &  \\ \hline
\end{tabular}%
}

\end{table}
\end{landscape}

The price-to-income gap coefficient consistently shows a negative impact on house price growth, particularly in the lower quantiles. In the short forecast horizon, at $\tau=0.1$ the price-to-income gap coefficient is negative for all models, while not being selected for the other quantiles.\footnote{The only exception is the Credit model, where it is selected for $\tau=0.5$ as well.} This findings align with \citet{mian2013political}, who suggest that financial constraints have an impact on marginal buyers in slower growth markets, whereas high-growth markets remain resilient on account of investor-driven demand. At $h=8$, the coefficients are showing significant effects and enter the models at all quantiles, the importance of the price-to-income gap becomes stronger as we increase the forecast horizon, except for $\tau=0.1$. This means that affordability constrains price growth over time, which is in line with the literature \citep{mayer2011housing}. The OLS results further confirm that affordability variables exert a more pronounced effect over longer forecast horizons, consistent with findings in the literature that elevated price-to-income ratios can presage downward adjustments in housing markets.

The results indicate that household income growth plays a limited role in explaining house price growth across most quantiles and models, with it only being selected for $h=8$. As such, there is a potential decoupling between income and house price dynamics, particularly in markets where price appreciation is driven by factors such as speculative demand or credit availability \citep{mian2013political}. Rising house prices may outpace income growth, reducing its relevance, especially in higher quantiles representing rapid growth scenarios. Additionally, the availability of government housing subsidies may weaken the direct link between income and price growth in the short run, as these subsidies enable households to participate in the market beyond what their income alone would permit. This effect is particularly pronounced in the upper quantiles, where housing benefits have a more pronounced effect, rendering incremental income changes less impactful. For $h=8$, the inverse is true: income is more important than housing loan subsidies in determining the shape of the distribution. Furthermore, at this horizon income has a relatively uniform impact across all quantiles, which is the reason why the OLS results largely mirror the quantile findings. This entails that income acts as a location shifter at $h=8$, with having no impact on the shape of the house price growth distribution.

In contrast, unemployment has a more pronounced and heterogeneous impact across the estimated quantiles. In particular, at the lower quantiles and the median ($\tau=0.1$, $\tau=0.3$, and $\tau=0.5$) unemployment has a negative impact at $h=4$ and $h=8$. This highlights that unemployment exerts its influence by shifting the left side of the distribution downward, making future house price growth more skewed. Rising unemployment increases the chance of sales while decreasing demand from consumers with limited resources.  Thus, price rise is suppressed.  The lagged nature of labour market dynamics on housing markets is highlighted by the fact that unemployment is not selected for any quantiles at $h=1$ for the Full model. These results are in line with \citet{riley2015house}, who examine the impact of unemployment on mobility preferences of low-income homeowners. The authors find that slower-growing market segments are more vulnerable to labour market shocks. The models point to both FISS and unemployment having a negative non-linear impact on the house-price growth distribution, but at different points in the the forecast horizon. Financial stress exerts its influence at shorter horizons and decreases as the horizon is increased, while unemployment's influence increases with the horizon. The OLS estimates confirm this pattern across the forecast horizons for FISS and unemployment. However, because OLS averages effects over the whole distribution, the projected coefficient are less negative than those seen at the lower quantiles in the quantile regressions. This suggests that although increased unemployment tends to slow price increases, it also makes the distribution of growth in home prices more skewed.

The impact of housing loans plus benefits is particularly notable in the upper quantiles ($\tau=0.7$ and $\tau=0.9$) of house price growth, where subsidies are associated with significant positive effects in all models that include the variable, reflecting their role in fuelling price growth.The largest influence of credit plus benefits variable on house price growth is estimated to be at $h=4$ and $\tau=0.9$. 

Lower quantiles not having housing credit plus benefits selected, while it is always present for the upper quantiles highlights that housing loan subsidies are `pro-cyclical', leading to larger growth in house prices. This in turn, leads to an uneven distributional impact of housing loan subsidies on house price dynamics. This is in line with the critique of \citet{glaeser2017extrapolative}, that such policies may contribute to overheating in already dynamic markets. In particular, it is likely that such housing benefits tend to increase prices in already high-growth markets while having limited influence on slower-growing or declining segments. This in turn can lead to further market segmentation, benefiting wealthier households or higher-value housing markets disproportionately. The combination of large negative coefficients for FISS at the lower quantiles and large positive credit effects in the upper quantiles suggests increasing dispersion and right-tail fuel from credit while left-tail risks are dominated by stress/unemployment.

On the supply side, the influence of building costs and construction permits on house price distribution can be seen in the full and housing supply focused model. In the Full model, construction permit coefficients become increasingly positive with higher quantiles at the short forecast horizon, but have a negative impact at the longest forecast horizon. This highlights how issuing more permits helps reduce upward pressures on prices only in the long run, as it takes time for these new developments to appear on the market as housing. 

When building costs are also included in the Supply model a more nuanced picture emerges about supply sides impact on the house price dynamics' distribution. In particular, we can see that building costs have a positive coefficient for all forecast horizons, and most quantiles (except $\tau=0.1$). The positive coefficient indicates that higher construction costs translate into elevated house prices, suggesting that developers pass the bulk of incurred costs on to buyers. This suggests that supply-side constraints not only affect the overall price level but also distort the price distribution by amplifying price growth in markets with tight supply conditions having stronger impact on the upper quantiles of the house price growth distribution. This is in line with the literature on housing supply \citep{glaeser2005urban}. In particular, housing market has a kinked supply curve: it is very responsive when prices meet or exceed construction costs but much less flexible when prices fall below that level. Furthermore, existing homes, being durable goods, remain on the market for a long time. As a result, in areas with slow or declining demand growth, house prices are limited by building costs.

Overall, there is a clear pattern for the different variables. In the short-run downside risk is dominated by financial stress, while unemployment and affordability is more important in the other forecast horizons at the lower quantiles. Upside risk is fuelled by credit (and subsidy) variable across the horizons. Housing supply variables reverse sign across horizons, and exerts a negative influence on house price growth, as housing permits materialise into new developments.

\subsection{Model fits}
\begin{table}[]
\caption{Out of sample results (smaller is better)}
\label{tab:OOSresults}
\centering
\begin{tabular}{lr|cccc}
\hline
 &  & CRPS & Centre & Left tail & Right tail \\ \hline
\multicolumn{2}{l|}{h=1Q} &  &  &  &  \\
 & Full & 0.8422 & 0.1681 & \textbf{0.2672} & 0.4070 \\
 & Supply & \textbf{0.8063} & \textbf{0.1591} & 0.2698 & \textbf{0.3774} \\
 & Credit & 0.8490 & 0.1701 & 0.2696 & 0.4093 \\ \hline
\multicolumn{2}{l|}{h=4Q} &  &  &  &  \\
 & Full & 0.5532 & 0.1129 & \textbf{0.1650} & 0.2753 \\
 & Supply & \textbf{0.5419} & \textbf{0.1101} & 0.1679 & \textbf{0.2640} \\
 & Credit & 0.5473 & 0.1109 & 0.1664 & 0.2700 \\ \hline
\multicolumn{2}{l|}{h=8Q} &  &  &  &  \\
 & Full & \textbf{0.5010} & \textbf{0.0971} & \textbf{0.1464} & \textbf{0.2575} \\
 & Supply & 0.5283 & 0.1022 & 0.1579 & 0.2681 \\
 & Credit & 0.5517 & 0.1067 & 0.1571 & 0.2879 \\ \hline
\end{tabular}

\end{table}

To obtain the out-of-sample fits, the models are run with an initial sample size of 50, which is expanded 1 period at a time. The estimated betas are then used to fit the next out of sample period. To evaluate the forecasting performance of the different models, we use the quantile weighted CRPS (qwCRPS) from \citet{gneiting2011comparing} as our scoring rule. First, we compute the Quantile Score (QS), which is the (quantile) weighted residual associated with a forecasted observation, $\rho(\hat{y}_{t+h,p})$. The qwCRPS is then defined by integrating the QS over all quantiles:

\begin{equation}
    qwCRPS_{t+h} = \int^1_0 \; w_q QS_{t+h,q}dq,
\end{equation}

\noindent where $w_q$ represents the weighting function that focuses the evaluation on particular regions of the forecast distribution. This scoring rule is selected because different weighting schemes allow us to assess discrepancies in various parts of the distribution. In our analysis, we employ four distinct weighting schemes: $w_q^1=\frac{1}{Q}$, which assigns equal weight to every quantile\footnote{This is equivalent to taking the average of the weighted residuals for a given observation.}; $w_q^2=q(1-q)$, which emphasizes central quantiles; $w_q^3=(1-q)^2$, which gives greater importance to the left tail; and $w_q^4=q^2$, which accentuates the right tail. The out-of-sample results for the various weightings are shown in table (\ref{tab:OOSresults}).


The out-of-sample results reveal more than a ranking: they expose how different informational sets map to distinct forecast roles. At short and medium horizons the Supply specification produces the best overall and centre/right-tail performance ($h=1$: 0.8063 vs Full 0.8422 and Credit 0.8490; $h=4$: 0.5419 vs Full 0.5532 and Credit 0.5473; right tail $h=1$: Supply 0.3774 vs Full 0.4070). This is likely because supply-side indicators track market momentum and capacity constraints. In essence, permits move with market activity and can reflect near-term construction pipelines that affect transaction flows and short-term scarcity, while building costs immediately transmit input-side pressures into asking prices. In short horizons these variables capture much of the systematic variance in the central mass and the boom-side dynamics, while excluding extra macro-financial variables which may introduce additional estimation noise for short-run density forecasts.

By contrast, the Full model consistently outperforms on the left tail across horizons ($h=1$ left = 0.2672, $h=4$ left = 0.1650, $h=8$ left = 0.1464) and becomes the best overall model at the long horizon ($h=8$ overall = 0.5010 vs Supply 0.5283). The left-tail dominance is consistent with the coefficient evidence: financial stress displays large short-run negative effects and unemployment and affordability metrics gain importance as horizons lengthen. Mechanistically, stress and labour-market shocks alter credit supply, default risk and buyer participation in a non-linear way that disproportionately increases downside risk; these are the channels the Full model is explicitly designed to capture. Over longer horizons the accumulation and propagation of such macro-financial and affordability effects make broader macro sets — not just supply indicators — essential to model the full predictive distribution.

Both Full and Supply include key macro-financial predictors (e.g., FISS, unemployment, Price-to-Income gap), yet they differ in two policy-relevant ways: (1) the Full specification sometimes contains a credit/subsidy variable, and (2) the Supply specification includes building costs. The Full model’s advantage on the downside therefore appears to arise from (i) slightly different coefficient calibrations for stress and labour-market predictors, and (ii) the presence of the credit/subsidy indicator (and its joint interaction with stress/unemployment), which together help partial out the variables impact on the lower quantiles at $h=4$. As such the difference between the forecast performance at the lower tail reflect differences in coefficient magnitudes and selection patterns rather than an exclusive presence of certain variables in one model only.



%% file: Chapters/FinStab.tex
\subsection{Uncertainty decomposition over time}

While the coefficients (and the out-of-sample results) are important, they do not inform us about total impacts on the distribution. Specifically, how would these changes in the distribution translate to overall uncertainty regarding the housing market. To this end, we follow \citet{castelnuovo2025uncertainty} in creating a measure of uncertainty and skewness:

\begin{equation}
    U_t=\widehat{\mathcal{Q}}_{0.9}(y_{t+h}|x_t)-\widehat{\mathcal{Q}}_{0.1}(y_{t+h}|x_t)
\end{equation}

The reasoning for this measure of uncertainty is that, conditional on a predictive model, uncertainty can be approximated by the gap between chosen quantiles of the conditional distribution. When uncertainty is high, market participants’ expectations, represented here as a full probability distribution over future house-price growth, are more dispersed. As such economic agents assign non-negligible probability to a broad set of possible outcomes. By contrast, when uncertainty is low, probabilities assigned to extreme outcomes in the tails of the conditional distribution are smaller.

\begin{equation}
    S_t=\frac{[\widehat{\mathcal{Q}}_{0.9}(y_{t+h}|x_t)-\widehat{\mathcal{Q}}_{0.5}(y_{t+h}|x_t)]-[\widehat{\mathcal{Q}}_{0.5}(y_{t+h}|x_t)-\widehat{\mathcal{Q}}_{0.1}(y_{t+h}|x_t)]}{\widehat{\mathcal{Q}}_{0.9}(y_{t+h}|x_t)-\widehat{\mathcal{Q}}_{0.1}(y_{t+h}|x_t)}
\end{equation}

\begin{figure}[!h]
    \centering
    \begin{subfigure}[]{0.8\textwidth}
    \centering
        \includegraphics[width=\textwidth]{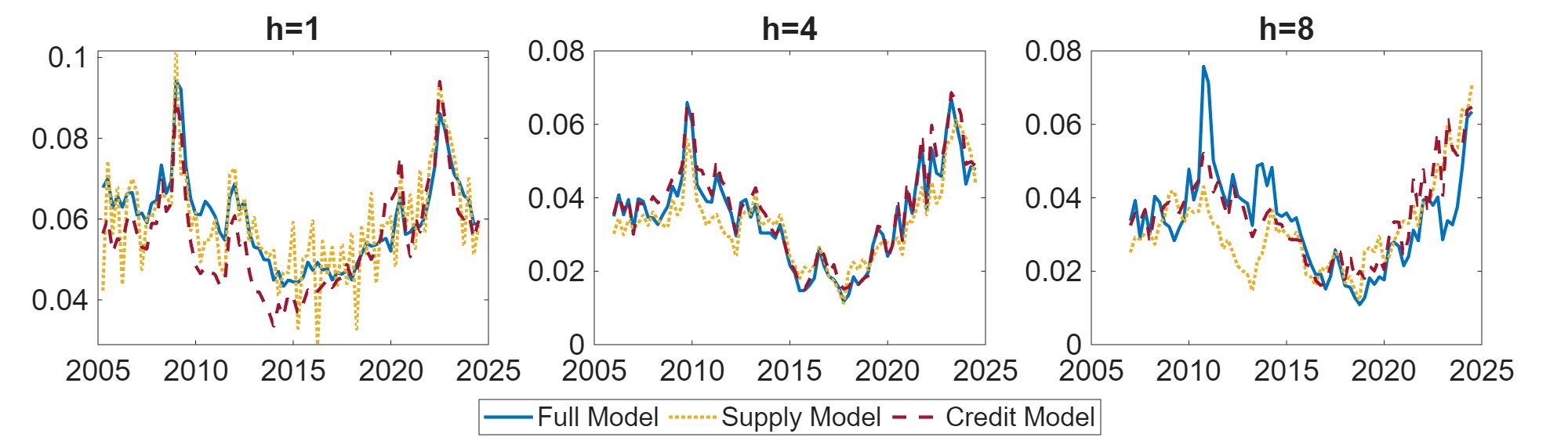}
        \caption{Uncertainty over time for the different models}
    \end{subfigure}
    \\
    \begin{subfigure}[]{0.8\textwidth}
    \centering
        \includegraphics[width=\textwidth]{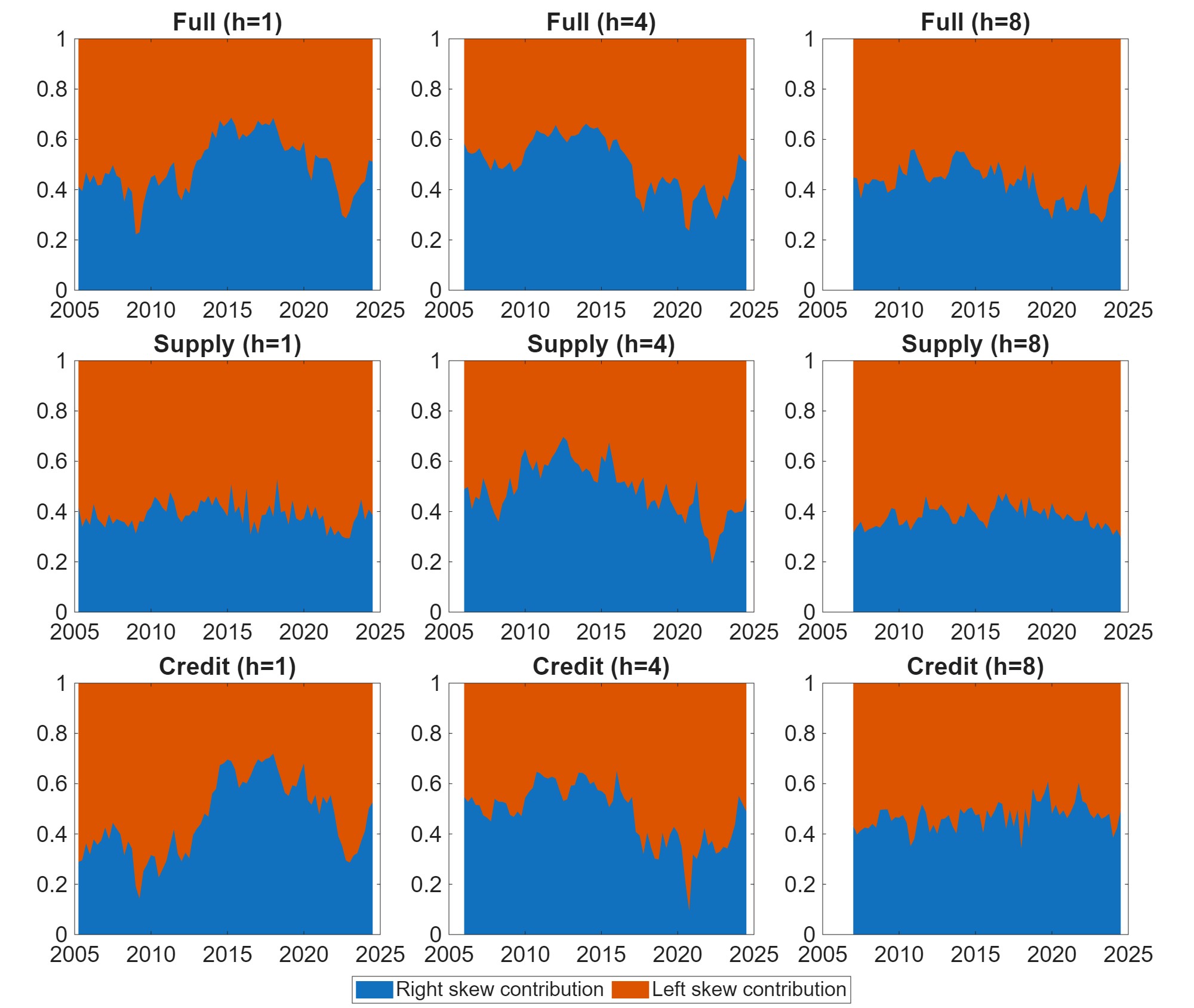}
        \caption{Left and Right side contribution of uncertainty}
    \end{subfigure}
    \caption{Uncertainty decomposition of the different models at the different horizons}
    \label{fig:UncertDecomp}
\end{figure}

The skewness metric provides a transparent decomposition of the share of conditional dispersion coming from the distribution’s left and right tails. Positive values of this measure indicate excess right side skew, while negative values indicate excess left side skew. Note that $S_t\in[-1,1]$, with 0 indicating no excess skew. On account of this, one can rescale $S_t$ and use it to decompose the $U_t$ into right and left tail contribution. This decomposition of uncertainty for the different models is shown in figure (\ref{fig:UncertDecomp}) along with the uncertainty measures for the different models.

The decomposition reveals distinct patterns in uncertainty drivers across forecast horizons and model specifications in the Hungarian housing market. For $h=1$, left-side skewness emerges as the primary driver of uncertainty for the Supply model, while for the Full and Credit models, left-side skew dominance emerges around the peaks of uncertainty. This is in line with the broader literature on asymmetric risk in financial markets: investors fear large negative shocks to their wealth, induced by rare crashes. Such large (negative) shocks to wealth are captured by the negative skewness of the distribution rather than volatility \citep{lemperiere2017risk}. It is worth noting that the supply model does not portray much temporal change in left vs right side skewness dominance, while the Credit and Full models track changes in the level of uncertainty driven by left sided skewness better. 

The uncertainty decomposition for $h=4$ and $h=8$ show that while right-side skewness becomes more prominent (in proportion to total uncertainty), left-side skewness maintains its importance throughout the sample period in the Hungarian market, especially around the times when total uncertainty peaks. These high uncertainty periods coincide with regimes of heightened market stress including the Global Financial Crisis (and the subsequent European debt crisis that significantly affected European economies) and the COVID period. Interestingly, the Full model captures the uncertainty peak around the Global Financial Crisis when looking at $h=8$, while the Supply model and Credit models fail to adequately reflect this period of housing market distress in Hungary as strongly. 

All models exhibit a notable spike in uncertainty around 2020, reflecting the market disruption associated with the COVID-19 pandemic. The importance of left skewness also increased in this period which is in line with the broader evidence that market downturns increase asymmetry in the transmission speed of negative tail risk information \citep{hu2025market}.

The persistent dominance of left-side skewness, especially when uncertainty peaks, across all forecast horizons suggests that downside housing market risks constitute a structural feature of the Hungarian market rather than a temporary phenomenon. This aligns with the theoretical work on real estate uncertainty, where financial risks reduce the demand for houses \citep{lee2025asymmetric} and with the broader literature on tail risk, where downside risk varies much more than upside risk during periods of tight financial conditions and high macroeconomic uncertainty \citep{huang2024financial}.

A key advantage of quantile regression applications to housing markets (such as \citet{zietz2008determinants} or \citet{mora2019determinants}), is that the method can identify coefficients across different quantiles while maintaining parameter stability. Stable estimation is important for policy applications as false signals lead to costly policy adjustments and degrades the authority's credibility. In this regard the Supply model's short-term uncertainty profile has a critical limitation: it displays excessive volatility from quarter to quarter. This extreme variability appears unrealistic for the Hungarian housing market context and suggests potential overfitting or model instability, raising concerns about the model's practical implementation for short-term policy applications by Hungarian authorities. 

Given this consistent left-tail dominance in driving uncertainty, we will focus exclusively on the Full model in the rest of the paper, as it has superior left tail forecast performance in Hungary.

\subsection{Impulse Response to Financial Risk build-up}

\begin{figure}
    \centering
    \includegraphics[width=\linewidth]{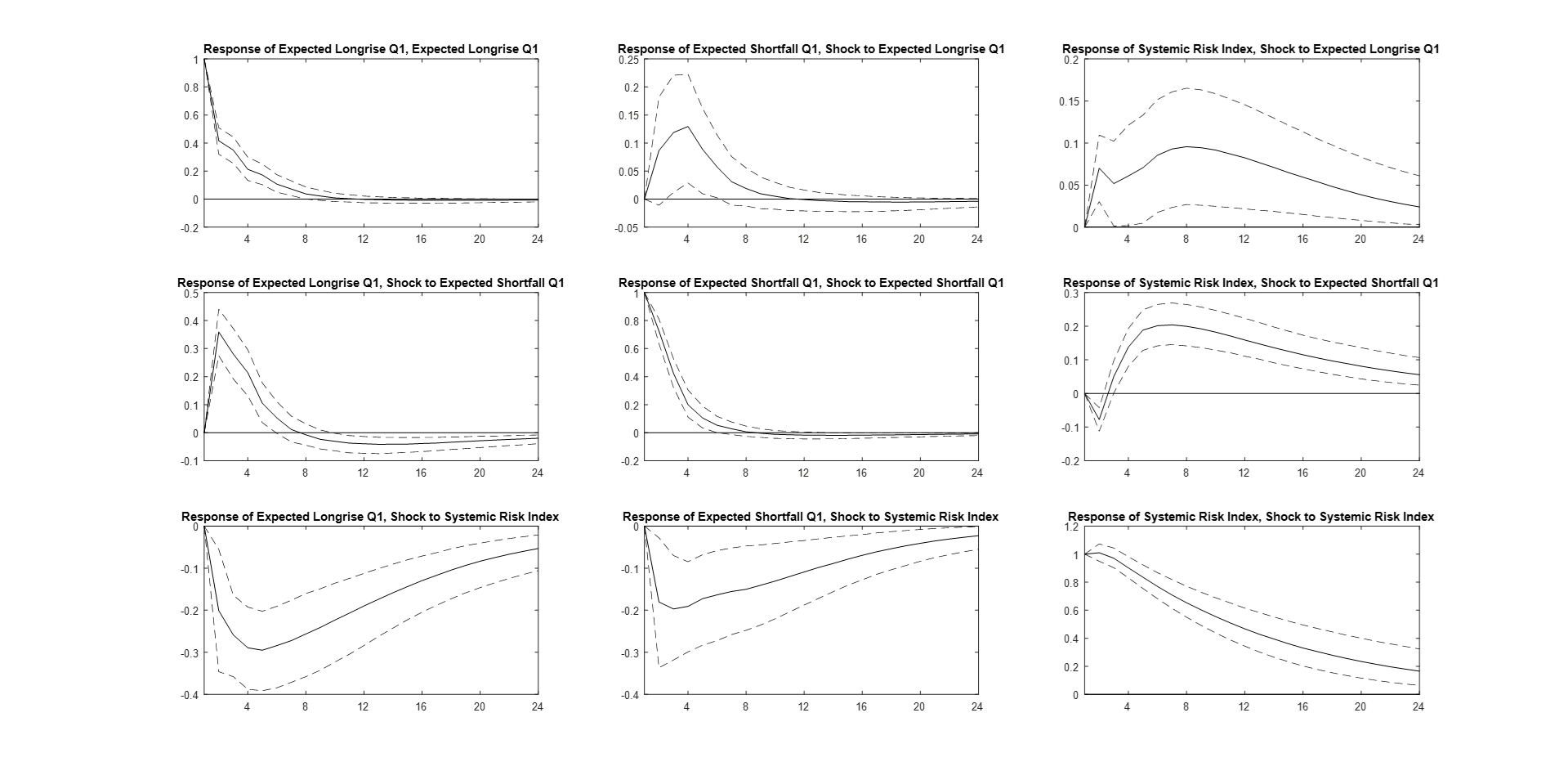}
    \caption{Using ES and EL of $h=1$ model}
    \label{fig:IRF_h1}
\end{figure}

\begin{figure}
    \centering
    \includegraphics[width=\linewidth]{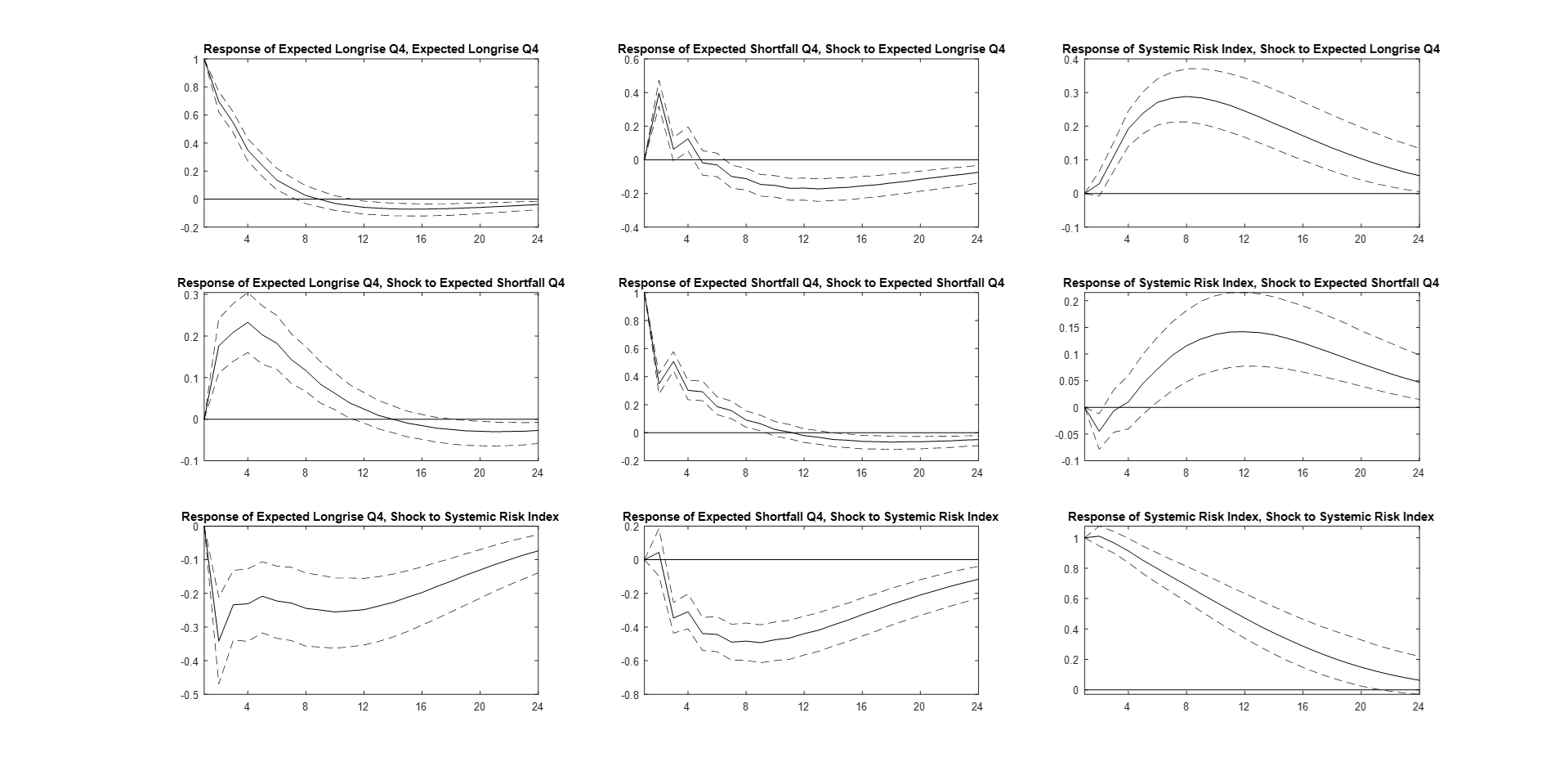}
    \caption{Using ES and EL of $h=4$ model}
    \label{fig:IRF_h4}
\end{figure}

In this section we will look at the financial stability implications of the House Price-at-Risk models. To do this we construct the Expected Shortfall (at $\tau=0.05$) (hereinafter ES) and Expected Longrise (at $\tau=0.95$) (hereinafter EL) from the fitted quantiles of the model using the method of \citet{mitchell2024constructing}. We will then construct a small VAR with Systemic Risk Index (SRI), which track the build up of financial booms. In this way we can track how upside and downside risk of house prices influence the overall financial market. Note that while FISS is a financial stress index, i.e. it tracks financial risks materialising, the SRI was built to track the build-up of financial risks over time. As such, the FISS and SRI measure different things. For more information see \citet{ESRB2021} and \citet{lang2019anticipating}. This analysis is similar to \citet{forni2025downside} and \citet{castelnuovo2025uncertainty}, but instead of focusing on how uncertainty and skewness changes impacts the economy, we model how the tails of the distribution influence the build-up of financial risks. The impulse response functions of shocking expected shortfall and expected longrise are shown in figures (\ref{fig:IRF_h1}) and (\ref{fig:IRF_h4}), for $h=1$ and $h=4$ respectively. Note, that all shocks are positive in the figures, i.e. we test the impact of moving parts of the house price growth distribution up. Because of the linearity of VAR models, if one is interested in the impact of worsening house price dynamics, the IRF of expected shortfall and expected longrise shocks can be flipped.

The impulse response analysis reveals asymmetries in how upside and downside house price risks propagate through the Hungarian financial system. For $h=1$ an increase in EL generates a delayed response in ES after four quarters. As such the whole distribution shifts up in the $4^{th}$ quarter, but for rest of horizons, there is increased skewness induced, as the upper part of the housing distribution rises for longer. A similar pattern can be seen when ES is shocked, with the difference that EL portrays a stronger and more persistent response, lasting approximately 1.5 years. This suggests that downside housing market risks have more substantial and enduring impacts on the overall distribution than upside risks. 

The figures also show a significant and persistant impact on the financial system from the housing market. This aligns with research demonstrating that housing booms and busts have quite often been detrimental to both financial stability and the real economy \citep{jorda2015leveraged,baek2024real}. EL shocks consistently increase the SRI over the impulse response horizon, with the magnitude of this effect amplifying when the shock persists over longer periods (captured through $h=4$ shocks). In contrast, ES shocks initially reduce systemic risk, but this effect reverses after 3 quarters. ES shocks having larger impacts on systemic risk (than EL shocks) suggests that downside housing market developments pose greater threats to financial stability. 

When systemic risk itself increases, both EL and ES decline, with EL showing more persistent negative responses. This pattern is indicative of elevated financial risk constraining the upside potential of housing markets. This suggests that systemic risk build-up creates asymmetric constraints on housing market dynamics. The IRF on $h=4$ reveals that sustained EL increases lead to greater dispersion in housing price distributions, with ES initially rising but subsequently declining, indicating that prolonged upside housing market pressure eventually generates complex distributional effects. This asymmetric transmission mechanism, that shocks to the financial system have longer impact on the housing market (as captured by Expected Shortrun and EL) is consistent with broader evidence from financial markets: cross-market risk spillover features an asymmetric pattern, with housing markets receiving more information from the financial system \citep{cunningham2006house}.

\subsection{Spillovers across sectors}

The financial stability view of housing market necessitates understanding how developments of the tails of the housing market influence the financial system. As \citet{elhorst2021cross} highlights, spillovers are not limited to the transmission of shocks across geographical borders: they extend to interconnected markets, where disturbances in one market can cascade into others. As such, one can use the measure of \citet{diebold2009measuring,diebold2012better,diebold2014network}, originally developed to capture spillovers geographically, to capture the spillovers of the tails of the house price distribution to and from the financial system, as captured by SRI. 

The core idea of the connectedness approach of \citet{diebold2009measuring} is to quantify how shocks in one variable affect others, offering a view of transmission channels. The approach is based on forecast error variance decomposition (FEVD) from vector autoregressive (VAR) models, allowing one to compute how much of the forecast error variance in one variable is attributable to shocks in other variables. As such we can take the VAR from the previous section and decompose the forecast error variance of each variable. These forecast error variance decompositions measures how much of the forecast error variance of each variable is explained by shocks to itself (own variance) and by shocks to other variables in the system (spillover variance). For a given $h$-period-ahead forecast error variance of variable $i$, the total variance contribution from shocks to variable $j$ (including $i$) is denoted as $\theta_{ij}(h)$. This measure is the percentage of forecast error variance of variable $i$ that is due to shocks in variable $j$, for $j = 1, \dots, N$, where $N$ is the total number of variables in the VAR. The FEVD is computed using the following equation:

\begin{equation}
    \theta_{ij}(H) = \frac{\sum_{h=1}^{H} \left( e_i' \Phi_h \Sigma e_j \right)^2}{\sum_{h=1}^{H} \left( e_i' \Phi_h \Sigma \Phi_h' e_i \right)}    
\end{equation}

\noindent where $\Phi_h$ is the moving average coefficient matrix derived from the VAR, $\Sigma$ is the covariance matrix of the error terms, and $e_i$ and $e_j$ are selection vectors.

\begin{table}[t]
\centering
\caption{Pairwise connectedness measures}
\label{tab:ConnectMeas}
\begin{tabular}{l|cc}
\hline 
 & $h=1$ & $h=4$ \\ \hline
$C_{ES,SRI}$ & 2.625 & 2.669 \\
$C_{EL,SRI}$ & 2.232 & -0.818 \\ \hline
\end{tabular}
\end{table}

In the spillover analysis we follow \citet{diebold2012better} in using the generalized impulse response function (GIRF) approach proposed by \citet{pesaran1998generalized} which is invariant to the ordering of variables. By eliminating the need for an arbitrary variable ordering, the GIRF allows for a more robust estimation of spillovers and connectedness without the need for testing various ordering. This is particularly useful in our setting as it is not clear what would be the best order for the ES and EL of the housing market.

A known drawback of the GIRF approach is that, unlike the Cholesky decomposition, the FEVD often does not sum to one. To correct this, we follow the common practice of normalizing the contributions by dividing each element by the corresponding row sum.\footnote{This issue arises because the generalized impulse response function does not impose the same orthogonality condition as the Cholesky decomposition. To resolve the problem, the individual variance contributions from each variable are typically divided by the sum of contributions in each row, ensuring that the FEVD sums to 1.} To estimate the different connectedness measures we use the publicly available MATLAB implementation of \citet{pham2024diebold}.

Using the FEVD spillover matrix we can construct pairwise directional connectedness between SRI and ES (or EL). This index will measure the direction of connectedness between SRI and ES (or EL). These are calculated as $C_{i,SRI}=Spillover_{SRI \leftarrow i}-Spillover_{i \leftarrow SRI}$. As such, when the index is positive, then ES (or EL) impacts SRI more, and if it is negative then SRI influences ES (or EL) more. These connectedness indices are presented in table (\ref{tab:ConnectMeas}).

The table reveals distinct characteristics across forecast horizons with the longer-term model demonstrating more spillovers from the SRI compared to the short-term specification. This indicates that financial conditions have more pronounced and persistent effects on housing market tail risks over extended periods. Specifically we find that in the longer horizon, SRI has larger impact on EL than the other way round. Interestingly, this is not the case for ES, where ES impacts SRI more regardless of which horizon is considered. These findings are consistent with the earlier finding that left-side skewness dominance persists across all forecast horizons, suggesting that downside housing market risks maintain structural importance regardless of the time frame considered. ES shows more spillover compared to EL, reflecting the asymmetric nature of risk transmission that mirrors the persistent left-tail dominance observed in the uncertainty decomposition analysis.

\begin{figure}[t]
     \centering
     \begin{subfigure}[b]{\textwidth}
         \centering
    \includegraphics[width=\linewidth]{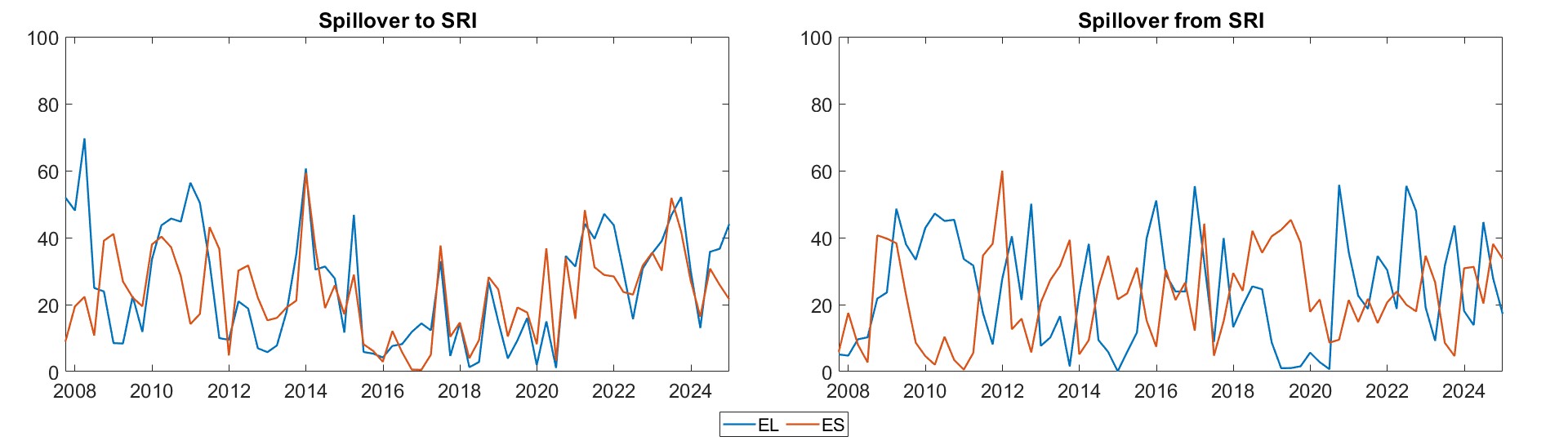}
    \caption{Spillovers for $h=1$}
    \label{fig:SpillQ1}
     \end{subfigure}
     \vfill
     \begin{subfigure}[b]{\textwidth}
         \centering
    \includegraphics[width=\linewidth]{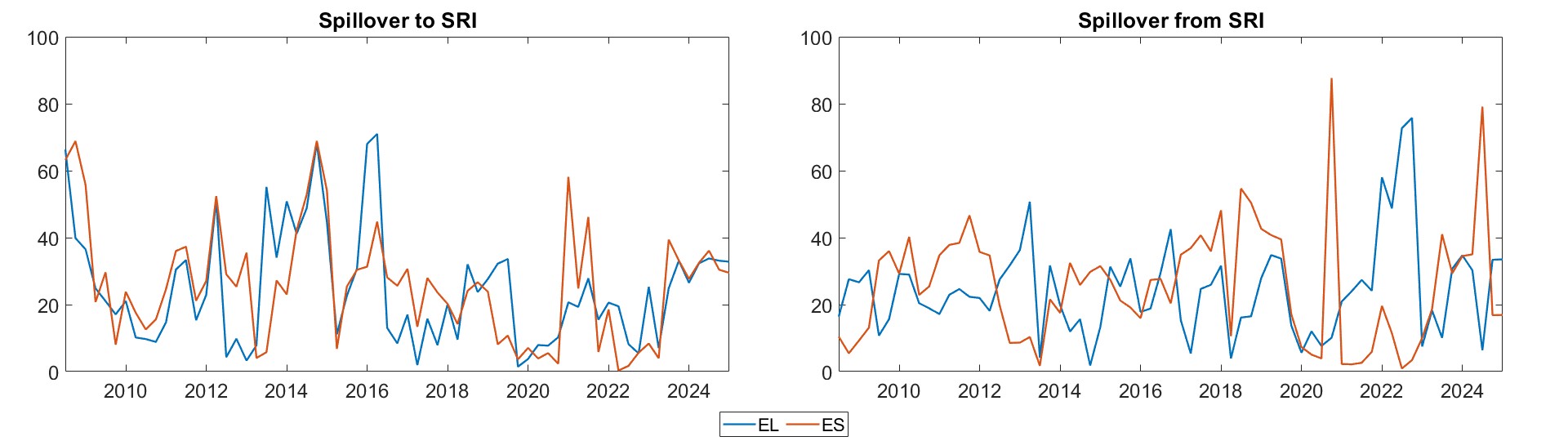}
    \caption{Spillovers for $h=4$}
    \label{fig:SpillQ4}
     \end{subfigure}
     \caption{Diebold-Yilmaz spillover over time}
\end{figure}

The Diebold-Yilmaz spillover method also allows for the construction of a time-varying spillover measure. This dynamic approach enables the ability to gauge when asymmetric house price dynamics impacted the financial system (and vice versa). Calculating the time-varying connectedness measure is done by applying a rolling window estimation approach to the underlying VAR model, where the model parameters and FEVD are recalculated over successive overlapping sub-samples of fixed length.\footnote{We fix the estimation window length to be 10 quarters long.} 
The results of the time varying spillover measure is shown in figures (\ref{fig:SpillQ1}) and (\ref{fig:SpillQ4}) for the $h=1$ and $h=4$ model respectively.

The first thing to note is that there are substantial bi-directional transmission effects between housing market tail risks and financial stability in Hungary, providing empirical validation for the theoretical relationships identified in the impulse response analysis. The figures show that housing market extremes not only influence financial stability but are equally affected by financial conditions. This bi-directional relationship aligns with established research \citep{jorda2015leveraged,baek2024real} while simultaneously confirming that systemic financial stress constrains housing market dynamics through reverse transmission channels.


The figures also show high spillover values around the 2008 GFC, where both ES and EL spillovers to the SRI approached 70–75 percent. The rise in spillovers from ES reflects heightened sensitivity of financial institutions to changes in extreme negative outcome probabilities of the housing market \citep{duan2025housing}. The sustained elevation in spillover measures of ES (and EL for the longer horizon) during crisis periods, such as the GFC and the COVID era, is consistent with the persistent IRF dynamics documented earlier, where housing market shocks propagate gradually through financial stability channels rather than dissipating quickly.


The finding of ES impacting SRI during the 2008 GFC is particularly compelling in the Hungarian context, as the country experienced an extensive and prolonged financial crisis with severe social consequences stemming primarily from foreign currency mortgage loans used for housing purchases. The substantial spillover transmission observed during this period demonstrates the HaR's relevance for policy applications. Hungarian authorities should also monitor these spillover measures as complementary indicators of emerging systemic risk.

%% file: Chapters/Policy.tex
Upper quantiles of house‐price risk (HaR) provide valuable signals for maintaining financial stability. When these upper quantiles rise, it typically indicates that housing demand is growing more rapidly than supply, which can in turn stimulate further credit demand. One promising regulatory approach is to use the gap between the upper quantiles and the median house price as a benchmark to adjust loan‐to‐value (LTV) requirements. In essence, borrower‐based regulation can be informed by HaR measures to help contain credit growth that is driven by escalating house prices. Research has shown that supply constraints amplify house‐price increases when demand surges occur \citep{hilber2016impact}, reinforcing the case for dynamic LTV calibration. 

Furthermore, borrower-based regulation can not only be dynamic, but spatially differentiated taking into account regional differences. \citet{hartmann2015real} emphasises the need for region-specific regulatory measures given the geographically differentiated nature of property markets, suggesting that such policies are instrumental in countering potential overvaluation. In essence, housing market segments that are not being speculated on should not be penalised.

Our analysis also reveals that housing subsidies tend to exert a strong upward pressure on the upper quantiles of house prices, especially in short and medium forecast horizon. This effect suggests that such subsidies can be pro‐cyclical in the sense that it increases the likelihood of further house‐price growth. Recent evidence from the Irish housing market supports the view that pro‐cyclical housing policies can intensify boom‐bust cycles \citep{norris2017tale}. If HaR is incorporated into macroprudential frameworks, the effect of these subsidies  can be taken into account more effectively and policymakers can calibrate borrower-based regulation to offset their pro‐cyclical influence. In other words, although mortgage subsidies are designed to enhance housing affordability and borrower-based tools' limits aim to preserve financial stability, their simultaneous use may inadvertently counteract one another. For welfare maximization, it is therefore critical for policymakers to calibrate subsidies and borrower-based tools in a targeted way to boost affordability without undermining stability.

%% file: Chapters/Conclusion.tex
The focus of this paper is to investigate the impact of credit and housing policies on the Hungarian housing market through a Value-at-Risk lens. This approach enables us to explore if these types of policies have a nonlinear impact on house price dynamics. Of particular interest is the impact of subsidies on the distribution of house price growth. We estimated the HaR model across three specifications and multiple horizons using quantile regression with adaptive LASSO variable selection following the methodology of \citet{szendrei2023revisiting}.

Financial stress quantified by the Factor-based Index of Systemic Stress (FISS) has a sizable influence on short- and medium-term house price risks. Its coefficient has the "usual" slope experienced in Growth-at-Risk models; hence it effects negatively house price growth with more pronounced impact on the lower quantiles. The misalignment indicator is primarily important over 1 and 2-year time horizons and turned to be one of the most influential regressors. Macro variables have a significant impact over the long-term, unemployment effects negatively house price growth at 1 and 2-years horizon, while increasing income has a positive impact only in a longer run, 2-years. The new credit variable supplemented with subsidies has a greater impact on housing prices than new credit itself and has a substantial effect in the upper quantiles. Credit indicators lose significance over some time horizons when supply-side drivers are included. 

Supply side factors also exert an influence on the house price growth distribution. Building costs have a positive effect across forecast horizons and most quantiles, indicating that higher construction costs are largely passed on to buyers. Construction permits act as a demand proxy in the 1 and 4-quarter periods, so have a positive impact, while in the longer term, the construction of new flats can have a moderating effect on house prices through increased supply. This variable has a greater impact on the right tail of the distribution.

Beyond coefficient estimates, we examined uncertainty and skewness decomposition to understand overall market dynamics. To do this, we followed \citet{castelnuovo2025uncertainty} in creating a measure of uncertainty and skewness. The decomposition revealed distinct patterns in uncertainty drivers for the Hungarian housing market, for h = 1, left-side skewness emerged as the primary driver of uncertainty for both the Full and Supply models. This experience is in line with the literature on asymmetric risk in financial markets: investors fear large negative shocks to their wealth, induced by rare crashes. The uncertainty decomposition for h = 4 and h = 8 showed that while right-side skewness became more prominent, left-side skewness maintained its importance throughout the sample period. All models exhibited a notable spike in uncertainty around 2020, reflecting the market disruption associated with the COVID-19 pandemic. The importance of left skewness also increased in this period which is in line with the broader evidence that market downturns increase asymmetry in the transmission speed of negative tail risk information.

Our analysis of financial stability implications constructs Expected Shortfall and Expected Longrise from the fitted quantiles using the method of \citet{mitchell2024constructing}. These measures are integrated into a vector autoregression with the Systemic Risk Index to capture how housing market tail risks influence broader financial conditions. The impulse response analysis reveals pronounced asymmetries in how housing market tail risks propagate through the financial system. Expected Longrise shocks persistently increase systemic risk, with effects amplifying over longer periods. By contrast, Expected Shortfall shocks initially reduce systemic risk but reverse after three quarters, generating substantially larger overall impacts than upside shocks. This pattern suggests that downside housing developments pose greater threats to financial stability than equivalent upside movements. When systemic risk itself increases, both Expected Longrise and Expected Shortfall decline, with Expected Longrise exhibiting more persistent negative responses. This indicates that elevated financial risk constrains housing market upside potential while amplifying downside risks, creating asymmetric feedback dynamics. The persistent impact of financial system shocks on housing market tail risks aligns with broader evidence that housing booms and busts significantly affect financial stability and the real economy, and reflects the documented pattern whereby housing markets receive more information from the financial system than vice versa \citep{jorda2015leveraged,baek2024real,cunningham2006house}.

The financial stability view of housing market necessitates understanding how developments of the tails of the housing market influence the financial system. Spillovers are not limited to the transmission of shocks across geographical borders: they extend to interconnected markets. We use the methodology of \citet{diebold2009measuring}, to capture the spillovers of the tails of the house price distribution to and from the financial system, as captured by SRI. The method finds substantial bi-directional transmission effects between housing market tail risks and financial stability in Hungary. We also found that financial conditions have more pronounced and persistent effects on housing market tail risks over extended periods. This horizon-dependent behaviour is consistent with the earlier finding that downside housing market risks maintain structural importance regardless of the time frame considered. The spillover transmission observed during Global Financial Crisis period demonstrates the HaR’s relevance for policy applications.

These results suggest that HaR is relevant from a financial stability point of view and could be a valuable quantitative element of the Macroprudential Toolkit. For example, borrower-based regulation can be informed by HaR measures to help contain credit growth that is driven by escalating house prices. 